\documentclass[fleqn,usenatbib]{mnras}
\usepackage[T1]{fontenc}
\usepackage{graphicx}   
\usepackage{amsmath}    
\usepackage{amssymb}    
\usepackage{float}
\usepackage[dvipsnames]{xcolor}

\newcommand\cmtvb[1]{{\color{red}[VB: #1]}}
\newcommand\cmtlg[1]{{\color{magenta}[LG: #1]}}
\newcommand\cmtnee[1]{{\color{ForestGreen}[NEE: #1]}}

\title[M87 black hole mass and spin]{M87 black hole mass and spin estimate through the position of the jet boundary shape break}

\author[Nokhrina et al.]{\parbox{\textwidth}{
E.~E.~Nokhrina$^{1}$\thanks{E-mail: nokhrina@phystech.edu},
L.~I.~Gurvits$^{2,3}$,
V.~S.~Beskin$^{1,4}$,
M.~Nakamura$^{5}$,
K.~Asada$^{5}$,
K.~Hada$^{6,7}$
}
\vspace{0.4cm}\\
\parbox{\textwidth}{
$^1$Moscow Institute of Physics and Technology, Dolgoprudny, Institutsky per., 9, Moscow region, 141700, Russia\\
$^2$Joint Institute for VLBI ERIC, Oude Hoogevceensedijk 4, 7991 PD Dwingeloo, The Netherlands \\
$^3$Department of Astrodynamics and Space Missions, Delft University of Technology, Kluyverweg 1, 2629 HS Delft, The Netherlands \\
$^4$Lebedev Physical Institute, Leninsky prosp.~53, Moscow, 119991, Russia \\
$^5$ Institute of Astronomy \& Astrophysics, Academia Sinica, 11F of Astronomy-Mathematics Building, AS/NTU No. 1, Taipei 10617, Taiwan\\
$^6$Mizusawa VLBI Observatory, National Astronomical Observatory of Japan, 2-12 Hoshigaoka, Mizusawa, Oshu, Iwate 023-0861, Japan\\
$^7$Department of Astronomical Science, The Graduate University for Advanced Studies (SOKENDAI), 2-21-1 Osawa, Mitaka, Tokyo 181-8588, Japan\\}
}

\begin{document}

\date{Accepted \dots Received \dots; in original form \dots}

\pagerange{\pageref{firstpage}--\pageref{lastpage}} \pubyear{}

\maketitle

\label{firstpage}

\begin{abstract}
  We propose a new method of estimating a mass of a super massive black hole residing in the center of an active galaxy. The active galaxy M87 offers a convenient test case for the method
  due to the existence of a large amount of observational data on the jet and ambient environment properties in the central area of the object. We suggest that the observed transition of a jet boundary shape from a parabolic to a conical form is associated with the flow transiting from the magnetically dominated regime to the energy equipartition between plasma bulk motion and magnetic field. By coupling the unique set of observations available for the jet kinematics, environment and boundary profile with our MHD modelling under assumption on the presence of a dynamically important magnetic field in the M87 jet, we estimate the central black hole mass and spin. The method leads us to believe that the M87 super massive black hole has a mass somewhat larger than typically accepted so far. 
\end{abstract}


\section{Introduction}
\label{s:intro}

The object Messier 87 (also known as NGC\,4486 and Virgo~A; hereafter -- M87 in short) is a super-giant elliptical galaxy. At the redshift of $z=0.0043$\footnote{NASA/IPAC Extragalactic Database, http://ned.ipac.caltech.edu, accessed 2019.03.14}, M87 is one the closest galaxies with active galactic nuclei (AGN). Long before identification as an AGN, the object attracted attention as the first jet, discovered in optical observations a century ago \citep{Curtis1918}. This jet, later detected in radio emission, has become a test bench for major models of AGN phenomena. Together with the Crab Nebula, M87 was one of the first celestial objects which facilitated the role of synchrotron emission in astrophysics \citep{ISSh58}.

The bright radio jet in M87 is one of extragalactic structures with best studied morphological properties on the angular scales from arcminutes down to sub-milliarcseconds. The external medium in the inner area of M87 is also best studied among AGN of various classes. M87 is the only galaxy with measurements of particle number density and a temperature of ambient medium at the distance to the central source $\sim 10^5$ gravitational radii, which is very close to a Bondi radius \citep{DiMat-03, RF15}. There is an extensive information on kinematics and jet transversal structure (see, e.g., \citealt{Mertens16}, \citet{Asada12}, \citet{Nakamura+18}, \citet{Hada18}, \citet{Lister-19}). All these observing data make the jet in M87 an ideal object for application of the theoretical models which connect the physical properties of the jet and its ambient medium.

A change in M87 jet shape along its extension has been first reported by \citet{Asada12}. It was shown that the power index $k$ in the dependence of jet width $d \propto r^{k}$ on the deprojected distance $r$ from the ``central engine'' along the jet changes at a $r \sim 100$ pc from $k \approx 0.6$ at small distances to $k \approx 0.9$ at large ones. Later the same ``cabing'' jet boundary shape behaviour was discovered for 1H0323+342 by~\citet{Hada18}, and the position of a break in this source suggested that the mass of a central black hole in 1H0323+342 might be underestimated~\citep{Hada18}. The jet geometry transition was also reported for NGC 6251 \citep{Tseng16}, for NGC 4261 \citep{Nakahara18}, and for Cyg-A \citep{Nakahara19}. As demontstrated recently by \citet{Kov-19}, a similar morphological pattern in jet shape (``cabing point'') is observed in ten nearby AGN.

The modern AGN paradigm associates many of their manifestations with the presence of a super-massive black hole (SMBH) as the major galactic gravitator. The SMBH mass defines the appearance of AGN and their major observable characteristics. Currently available estimates of the SMBH mass in M87 are based on a variety of measurements and corresponding interpretations. Over the past two decades these values were reported within the range from $M = (3.2\pm 0.9)\times 10^{9}M_{\odot}$ to $M = 9.5^{+0.22}_{-0.23}\times 10^{9}M_{\odot}$ \citep{Macchetto+97, Gebh-Thomas09, Geb11, Walsh+13, Oldham+16} based on 
the dynamical behaviour of various constituencies of galaxy population in the SMBH's gravitational field.

In this paper we propose a method of estimating BH mass for core-jet AGN that involves another SMBH manifestation -- a powerful relativistic jet launched from the circumnuclear area of the source. The method is based on MHD modelling \citep{BCKN-17} 
of a jet boundary shape and matching the model's ``cabing'' point in the jet shape with its observed position. 
We estimate the central BH mass and spin independently following the theoretical model by \citet{BCKN-17}, and using the measurements of jet parameters in M87: the ambient pressure, the plasma flow kinematics, jet opening angle, and the position of a jet shape break -- the ``cabing'' point. 

The paper is organized as follows. We describe the multifrequency observational data used to recover the M87 jet shape on the scales from $10^{-1}$~pc to $10^4$~pc and the ``cabing'' point position. In Section~3 we discuss briefly the MHD model that allows us to reconstruct the observed break in a jet shape for the smooth ambient pressure profile as well as all the model assumptions and the values needed to obtain the black hole mass and spin rate. In Section~4 we define the method of fitting the jet profile by two power-laws, and in Section~5 we discuss the errors. We present the results in Section~6.

Throughout the paper, we use the
$\Lambda$CDM cosmological model with $H_0=71$~km~s$^{-1}$~Mpc$^{-1}$, $\Omega_m=0.27$, and
$\Omega_\Lambda=0.73$ \citep{WMAP5_COSMOLOGY}.
 
\section{Observational data}
\label{s:data}

We use the multi-frequency radio interferometry data, reported by \citet{Asada12}, \citet{Hada13}, and \citet{Hada16}, and collected in the paper by \citet{Nakamura+18}. For each data set we use distance along the jet taking into account the error for the core data, and a radius of a detected feature with the error in radius determination (see \autoref{Trial}). The data sources and their thorough description are as follows. The 1.8~GHz data are obtained with MERLIN \citep{Asada12}. The 2.3, 5.0, 8.4, and 22~GHz data come from the Very Long Base Array (VLBA) as reported by \citet{Hada13}. The 15 and 43~GHz VLBA data have been reported by \citet{Asada12} and \citet{Hada13}. The 86~GHz data set is provided by the VLBA--High Sensitivity Array (HSA) observations \citet{Hada16}. \citet{Nakamura+18} use the luminosity distance $D_{\rm L}=16.7$~Mpc \citep{Blakeslee09}.

The VLBA core data at frequencies 5.0, 8.4, 15.4, 23.8, 43.2, 86.3~GHz are dedscribed by \citet{Hada13} and at frequencies 43 and 86~GHz -- by \citet{NA13, Hada16}. The Event Horizon Telescope (EHT) core data at 230~GHz obtained by \citet{Doeleman12, Akiyama15}. However, we do not use these core data to fit the jet boundary form in the parabolic domain due to large errors in the determination of the core position along the jet due to the core shift estimates \citep{Hada11}. We note however, that, described below, the major contribution in the BH mass estimate is provided by the jet boundary data on the scales larger than those of the core.

Detailed procedures of estimating the jet width are described in \citet{Asada12} and \cite{Hada13}. In short, we made transverse slices of the jet at various distances from the core. For each slice, we fitted a double-Gaussian function (if the slice is clearly resolved into a two-humped shape, which applies to most of the slices) or a single Gaussian (if the slice is single-peaked). We then defined the separation between the outer sides of the half-maximum points of the two Gaussians as the width of the jet at each distance (for the single Gaussian case, its deconvolved FWHM was taken as the jet width). Finally, the jet radius ($d$) at each distance was defined as a half of the jet width.


\section{Black hole mass and spin determination}

As was already stressed, the uniqueness of the M87 jet is in the availability of direct information not only on the jet boundary shape, but also on the ambient pressure  $P_{\rm ext}$ in close vicinity of the jet ``cabing'' region~\citep{Young02, DiMat-03, RF15}. Below we show that this additional information gives us the possibility to determine such key parameters of the ``central engine'' as the total magnetic flux $\Psi_0$ in the jet and the radius of the light cylinder $R_{\rm L} = c/\Omega$. In turn, magnetically arrested disk (MAD) assumption allows us to decouple the mass $M$ of the supermassive black hole and its spin parameter $a_*$. We designate the distance along the jet and the jet radius as $r$ and $d$, respectively. The function $d(r)$ determines the jet boundary shape. The position and radius of the ``cabing'' point at which the jet shape changes from parabolic to conical is designated as $r_{break}$ and $d_{break}$, respectively.

Below we use a model of the transversal structure of a jet based on the now generally accepted MHD theory within the framework of the approach of the Grad-Shafranov (GS) equation~\citep{HN89, PP92, Heyvaert96}. More precisely, on its one-dimensional cylindrical version, when a second-order partial differential GS equation can be reduced to two first-order ordinary differential equations~\citep{B87, Lery98, Lery99, BM00, BN09}. This approach has well proven itself for both non-relativistic and relativistic flows. In particular, just within this approach, it has been predicted theoretically that in a parabolic magnetic field, effective particle acceleration becomes possible~\citep{Beskin06}. Earlier, on the basis of solutions for quasi-spherical outflow, it was believed that effective acceleration in a magnetically dominated wind is impossible~\citep{Michel, KFO83, Bogovalov}. Later this conclusion has been repeatedly reproduced by numerical simulations (see, e.g.,~\citealt{McK06, N1}). Among other things, demonstrating the full consistency of a semi-analytical modelling with the numerical simulations, the existence of a denser core along a jet axis was obtained by \citet{BN09}. It was corroborated by independent numerical models~\citep{KBVK, TMN09, OFMV}. 
Finally, it was also shown by ~\citet{BZh13} how asymptotic relations obtained in the framework of the one-dimensional approach used in our present work make it possible to reproduce convincingly the results of numerical simulations for the black hole magnetosphere obtained by \citet{McKinney12}.

In what follows we use the most developed version, in which we assume that an electric current $J$ locked inside the jet \citep{BCKN-17}. In this model, the flow velocity and magnetic and electric fields vanish at the jet edge $d(r)$. In this case, the current sheet at the edge is absent. In numerical modelling, such a structure has been known for non-relativistic trans-sonic flows~\citep{Romanova}. Recently this structure was reproduced for relativistic outflows as well~\citep{BTch}.
The fall of a flow bulk motion Lorentz factor down to unity at the jet boundary is clearly seen in the numerical simulations by \citet{Nakamura+18}, in accordance with the assumption used here.

Assuming that the flow remains supersonic up to the very boundary of a jet, one can write down the force balance at the jet boundary as
\begin{equation}
\frac{{\rm d}}{{\rm d}r}\left(\frac{B_{\varphi}^2}{8\pi}+P\right) = 0.
\label{balance1}
\end{equation}
Here $B_{\varphi}$ is a toroidal magnetic field, which dominates the poloidal field $B_{\rm p}$ outside the light cylinder, and $P$ is a jet plasma pressure that transits smoothly into the pressure of the ambient environment. Integrating this equation through the thin boundary layer where the gradient of the gas pressure balances the magnetic stress, we obtain 
\begin{equation}
P_{\rm ext} = \frac{B_{\varphi}^2}{8 \pi}.
\label{balance2}
\end{equation}
Indeed, as was shown by \citet{Kov-19}, even for finite temperature the magnetic pressure dominates the force balance inside a jet up to the very thin boundary layer. The importance of a gas pressure at the jet boundary is supported by numerical simulations in \citep{Nakamura+18}.

Here we must emphasise the key difference of our model comparing to other ones (see, e.g.,~\citealt{Lyu09}). We note that in the framework of the approach considered here it is necessary to specify five integrals conserved on magnetic surfaces (energy density flux, angular momentum density flux, angular velocity of field lines, entropy, and mass-to-magnetic flux ratio). For the major part of the jet, we use standard values prescribed by the condition of a smooth crossing of the singular surfaces (Alfv\'enic and fast magnetosonic). However, near the outer boundary of the jet, the integrals were chosen in such a way that the condition of the total zero longitudinal current within the jet $J(d) = 0$ was satisfied. As was already stressed, such a structure of the integrals of motion corresponds to the results of numerical simulation~\citep{Romanova, BTch}.

Thus, the solution obtained by~\citet{BCKN-17} provides that the major part of an electric current is locked inside the bulk jet volume, and only a residual electric current $J_{\rm res}$, that defines $B_{\varphi}$ in \autoref{balance2}, is left in the outer thin jet layer. This implies that the characteristic toroidal magnetic field $B_{\varphi} = 2J_{\rm res}/c d$, which constitutes the major pressure at the jet edge, is much lower than in the models without an electric current closure.


For the cylindrical geometry, the Grad-Shafranov and Bernoulli equations, describing a full MHD flow, become a set of ordinary differential equations easily solvable numerically. The cylindrical flow solution reproduces accurately the axisymmetric flow solution if the derivatives along the jet are negligible. It was shown by~\citet{NBKZ15} that the solution, obtained within the  cylindrical geometry, is applicable for formation of the jet structure for the highly collimated flows. This allows us to use the cylindrical approach to the problem of modelling a well collimated jet.
In non-dimensional variables the solution of these equations depends only on the Michel's magnetization parameter $\sigma_{\rm M}$, which is defined as the ratio of Poynting flux to the plasma rest mass energy flux at the base of a flow.
Integrating the system of two ordinary differential equations describing internal structure of a jet (see~\citet{BCKN-17}
 for more details), we obtain a non-dimensional external pressure
\begin{equation}
\tilde{p}=\frac{P_{\rm ext}}{\left[\Psi_0/\left(2\pi R_{\rm L}^2\sigma_{\rm M}\right)\right]^2}
\label{p_def}
\end{equation} 
as a function of non-dimensional jet radius 
\begin{equation}
\tilde{d}=\frac{d}{R_{\rm L}}
\label{d_def}
\end{equation}
for different initial magnetizations $\sigma_{\rm M}$.
Here we use the natural inner scale for both poloidal and toroidal
magnetic field $B_{scale} = \Psi_0/(2\pi R_{\rm L}^2 \sigma_{\rm M})$, written through the total magnetic flux in a jet $\Psi_0$. The corresponding scale for pressure is $B_{scale}^2$. The integration of MHD equations for the given integrals (see~\citealt{BCKN-17}) provides the numerical factor, that relates this pressure scale with the corresponding jet inner pressure, needed to balance the ambient pressure.

Below we assume a power-law dependence of the ambient pressure on the distance from the central source:
\begin{equation}
P_{\rm ext}(r)= P_0\left(\frac{r}{r_0}\right)^{-b}.
\label{PBondi}
\end{equation}
Here $P_0$ is the ambient pressure amplitude at the distance $r_0$ from the BH. The exponent $b$ attains values between 1 and 2.5. The largest value 2.5 corresponds to the supersonic regime of a Bondi accretion of a gas described by the adiabatic equation of state $P\propto n^{\gamma}$ with $\gamma=5/3$. However, the recent theoretical studies of a gas accretion onto SMBH provide smaller values $b\in(1.0,\,2.1)$ \citep{QN00, NF11}, and the recent observations by \citet{Park-19} favour $b\lessapprox 2.0$. 
Thus, within our model we are able to determine the jet boundary shape $d(r)$ for the given ambient pressure profile $P_{\rm ext}(r)$. 

As was shown by~\citet{BCKN-17}, 
the obtained jet boundary dependence $d(r)$ has a pronounced break in the domain, where the flow transits from magnetically-dominated regime to the quasi-equipartition of plasma bulk motion kinetic energy density and the energy density of magnetic field. For the pressure profile predicted by the Bondi accretion model with $b\approx 2$, we obtain a clear transition from a parabolic to conical shape consistent with the results by~\citet{Asada12} and~\citet{Nakamura+18}. 


In our semi-analytical solution, the non-dimensional jet radius $d_*$ and ambient pressure $\tilde{p}_*$ are defined as functions of the Michel magnetization parameter $\sigma_{\rm M}$.  The results of the simulations are presented in \autoref{tableMod}. These simulations provide the position of the ``cabing'' point. The essence of our method is in comparing the jet's geometry at the ``cabing" point as obtained in the simulations with the observed shape of the jet.   

Using now \autoref{d_def} we obtain for the light cylinder radius
\begin{equation}
R_{\rm L}=\frac{d_{break}}{d_*(\sigma_{\rm M})}.
\label{RL}
\end{equation}
On the other hand, \autoref{PBondi}, rewritten for the ``cabing'' point, together with \autoref{p_def}, 
allows us to find the total magnetic flux $\Psi_0$ in a jet with the measured $r_{break}$:
\begin{equation}
r_{break}=r_0\left[\frac{\tilde{p}_*(\sigma_{\rm M})}{P_0}\left(\frac{\Psi_0}{2\pi R_{L}^2\sigma_{\rm M}}\right)^2\right]^{-1/b}. 
\label{rbreak}
\end{equation}
Here the pressure amplitude $P_0$ at the distance $r_0$ is known from the observations, while $\sigma_{\rm M}$ and $\tilde{p}_*$ --- from the modelling.


\begin{table}
  \caption{The non-dimensional parameters, which define the position of a ``cabing'' point, calculated for different magnetizations. The preferred values of $\sigma_{\rm M}$, basing on M87 kinematics, are 5, 10, and 20. \label{tableMod}}
 \begin{tabular}{ccc}
  \hline
  $\sigma_{\rm M}$ & $d_*$ & $\tilde{p}_*$ \\
   &  & ($10^{-5}$) \\
     (1) & (2) & (3) \\
\hline
5 & 33.6 & 1.39 \\
10 & 52.4 & 1.02 \\
20 & 79.8 & 0.75 \\
30 & 82.0 & 0.60 \\
40 & 115.9 & 0.59 \\
50 & 134.2 & 0.51 \\
\hline 
\end{tabular}
\end{table}

The results presented above are direct outcomes of MHD modelling of the jet structure \citep{BCKN-17}. For MHD models, the intrinsic length scale is the light cylinder radius $R_{\rm L}$, not the gravitational radius $r_{\rm g} = G M/c^2$. Thus, the position and radius of the cabing point depends on both the BH mass and its spin. Indeed, $R_{\rm L}$ and $r_g$ can be related 
for the maximum BH energy extraction rate condition $\Omega_{\rm F}=\Omega_{\rm H}/2$ \citep{BZ-77}. Here $\Omega_{\rm F}$ is a field lines rotational velocity, and $\Omega_{\rm H}$ is a BH angular velocity. Introducing the BH spin $a_*\in[0;\;1]$, we obtain the relation between $a_*$, $r_{\rm g}$ and $R_{\rm L}$:
\begin{equation}
a_*=\frac{8(r_{\rm g}/R_{\rm L})}{1+16\left(r_{\rm g}/R_{\rm L}\right)^2}.
\label{astar}
\end{equation}

Gravitational radius may be recovered if we assume that
the total magnetic flux $\Psi_0$ is locked with the mass accretion rate $\dot{M}$ \citep{NIA03}.  The numerical simulations by \citet{Tchekhovskoy_11} provide the following dependence: 
\begin{equation}
\Psi_0=\phi\sqrt{\dot{M}c}\,r_g,
\label{PsiMAD}
\end{equation}
with $\phi\sim 50$ in Gaussian units \citep{Tchekhovskoy_11} for a disk being in a magnetically arrested state (MAD). The same relation holds for a standart and normal evolution disk (SANE), with lower values of $\phi$ \citep{Narayan-12}. An analysis of a sample of 76 radio-loud sources \citep{ZCST14} gave the same result with $\phi=(52\pm 5)\Gamma\theta_j$, where
$\Gamma$ is a Lorentz factor of a bulk flow, and $\theta_j$ is a jet half-opening angle. For $\Gamma\theta_j\ll 1$ the disk state is SANE, not MAD (see discussion in \autoref{s:BHM}). 
For the Bondi accretion, we use the expression for an accretion rate that depends on the mass of a central BH. It is defined by the relation
\begin{equation}
\left(\frac{\dot{M}}{{\rm g/s}}\right)=C_{\dot{M}}\left(\frac{M}{10^9M_{\odot}}\right)^2,
\label{Mdot}
\end{equation}
where $C_{\dot{M}}$ depends on the ambient gas particle number density and temperature (see~\citet{DiMat-03} for more detail).
Substituting now \autoref{RL}, \autoref{PsiMAD}, and \autoref{Mdot} into \autoref{rbreak},
we finally obtain the following expression for the BH mass in M87:
\begin{equation}
\begin{array}{l}
\displaystyle\frac{M}{10^9M_{\odot}}=1.08\times 10^{2}\left(\frac{d_{break}/d_*(\sigma_{\rm M})}{\rm pc}\right)\left(\frac{r_{break}}{r_0}\right)^{-b/4}\times\\ \ \\
\displaystyle\sqrt{\frac{\sigma_{\rm M}}{\phi}}\left(\frac{P_0/\tilde{p}_*(\sigma_{\rm M})}{10^{-4}\,{\rm dyn/cm^2}}\right)^{1/4}\left(\frac{C_{\dot{M}}}{10^{24}\,{\rm g/s}}\right)^{-1/4}.
\end{array}
\label{MBH}
\end{equation}
We stress that we do not use in this formula
any results that were obtained under a priory assumption on the BH mass. Similarly, we use the expression \autoref{Mdot} and do not use a direct estimate of the accretion rate.

The described above method is based on an assumption that the real jet boundary, determined by the condition $\Psi=\Psi_0$, corresponds to the visible jet boundary. The latter is obtained as a cut at half maximum of intensity. Although in general case their coincidence might not be exact, it holds for our jet transversal structure model. We assume that the synchrotron self-absorbed emission is produced by highly relativistic plasma with an energy distribution $dn=k_{e}\Gamma^{-p}d\Gamma$. The emitting particle number density amplitude $k_{e}$ is either equal or proportional to the total local particle number density in a jet \citep{Lobanov98_coreshift,NBKZ15}. The intensity depends on the emission $\rho$ and absorption $\kappa$ coefficients for a synchrotron emission \citep{GS65}. They, in turn, are defined by the plasma conditions: the particle number density of emitting plasma and magnetic field amplitude. In case of an optically thick part of a flow, the intensity $I$ is depends on the magnetic field roughly as $\propto B^{-1/2}$, while in the optically thin region $I\propto n B^{(p+1)/2}$. In both cases, the profiles of $n$ and $B$ are such \citep{CBP19}, that the intensity grows towards the jet boundary, falling rapidly only in a very thin layer in its vicinity. As the flow is relativistic, the Doppler factor also affects the received intensity. For a high bulk Lorentz factor, the observer may be out of a cone of emission and receive the suppressed intensity, as can be seen in the Doppler maps by \citet{CBP19}. Thus, we expect that the observed jet boundary corresponds indeed to the model jet boundary $d(r)$. The effect of a jet slowing down at the boundary is expected in real jets and supported by the numerical simulations by \citet{McK06,Dexter12,Nakamura+18}.

\section{Break in the M87 jet shape}
\label{s:break_pos}

As was reported by \citet{Asada12}, the M87 jet boundary shape changes from approximately parabolic ($d\propto r^{0.5}$) to approximately conical ($d\propto r$). Our modelling predicts such the transition (the cabing point) as the flow accelerates from initially magnetically dominated regime to the energy equipartition. The change in a jet boundary shape occurs without a change in an ambient pressure profile. Thus, in order to compare theoretical predictions with observational data, we need to pin the observed position of the cabing point by approximating the jet boundary shape by two power laws to determine the SMBH mass. 

The procedure of fitting the two power laws is as follows.
We use the MERLIN and VLBA imaging data as described in \autoref{s:data} at frequencies 1.8, 2.3, 5.0, 8.4, 15.0, 22.0, 43.0, and 86.0 GHz \citep{Asada12, Hada13, Hada16, Nakamura+18}. For each frequency, we have a set of measured jet radii $d$,  de-projected (for the viewing angle of $14^\circ$) distances along the jet $r$, and an error in determination of $d$. Figure \ref{Trial} represents the $d(r)$ dependence for observational data their fit by two power laws. The first guess is that the change the power law index (cabing) occurs at the distance corresponding to the data obtained at 2.3~GHz. After this rough guess we divide the full sample into two sets (``parabolic'' and ``conical'') choosing a point from the 2.3~GHz sample as a boundary between them. For each such choice, we fit the power law parameters for two data sets of the full sample. We choose from the resulting set of possible approximations one that minimises the standard error in the expected conical domain. However, we also find the position of cabing point for every cut inside the 2.3~GHz sample, and use them to estimate the error in BH mass and spin values due to possible uncertainty in the cabing point determination. 

The obtained jet shape break for the ``best'' choice of a sample cut is at $r_{break}=43.41$~pc with the corresponding jet width radius $d_{break}=0.60$~pc. The power laws are:
$d_{\rm pc} = 0.07 \, r_{\rm pc}^{0.57}$ for the parabolic domain, and $d_{\rm pc} = 0.02 \, r_{\rm pc}^{0.90}$ for the conical domain. The result of this fit is presented in \autoref{Trial}. Here $d_{\rm pc}$ is a jet radius measured in~pc, and $r_{\rm pc}$ is a distance along a jet in~pc.

We use the full data at 1.8~GHz in contrast with \citet{Nakamura+18}, who excluded the farthest 4 points as a suspected jet wiggle that drives the conical domain fit to be more shallow. We have checked how deleting these points alter the results. We note that, indeed, the conical domain fit becomes steeper: $d\propto r^{0.92}$, with the cabing position moving to $r_{break}=45.16$~pc and $d_{break} = 0.62$~pc. However, this changes the final results for the mass and spin at the level $\sim 0.1\%$. In fact, this demonstrates the robustness of our result. The fit in the parabolic domain is very well defined. The final expression for the mass \autoref{MBH} has a term $d_{break}/r_{break}^{b/4}$, which varies very slowly as the jet boundary shape break follows the nearly parabolic trend (parabola holds, conical domain changes its slope). Because of this, we do not need to exclude the points that possibly reflect the local jet wiggle.


\section{Error budget}
\label{s:errors}

There are four major sources of errors in the mass determination by the method presented here: {\it (i)} errors from determining the cabing position; {\it (ii)} errors in the jet half-width; {\it (iii)} errors in $C_{\dot{M}}$ determination; {\it (iv)} errors due to uncertainty in $\Gamma\theta_j$ estimate.


\begin{figure*}
\centering
\includegraphics[
]{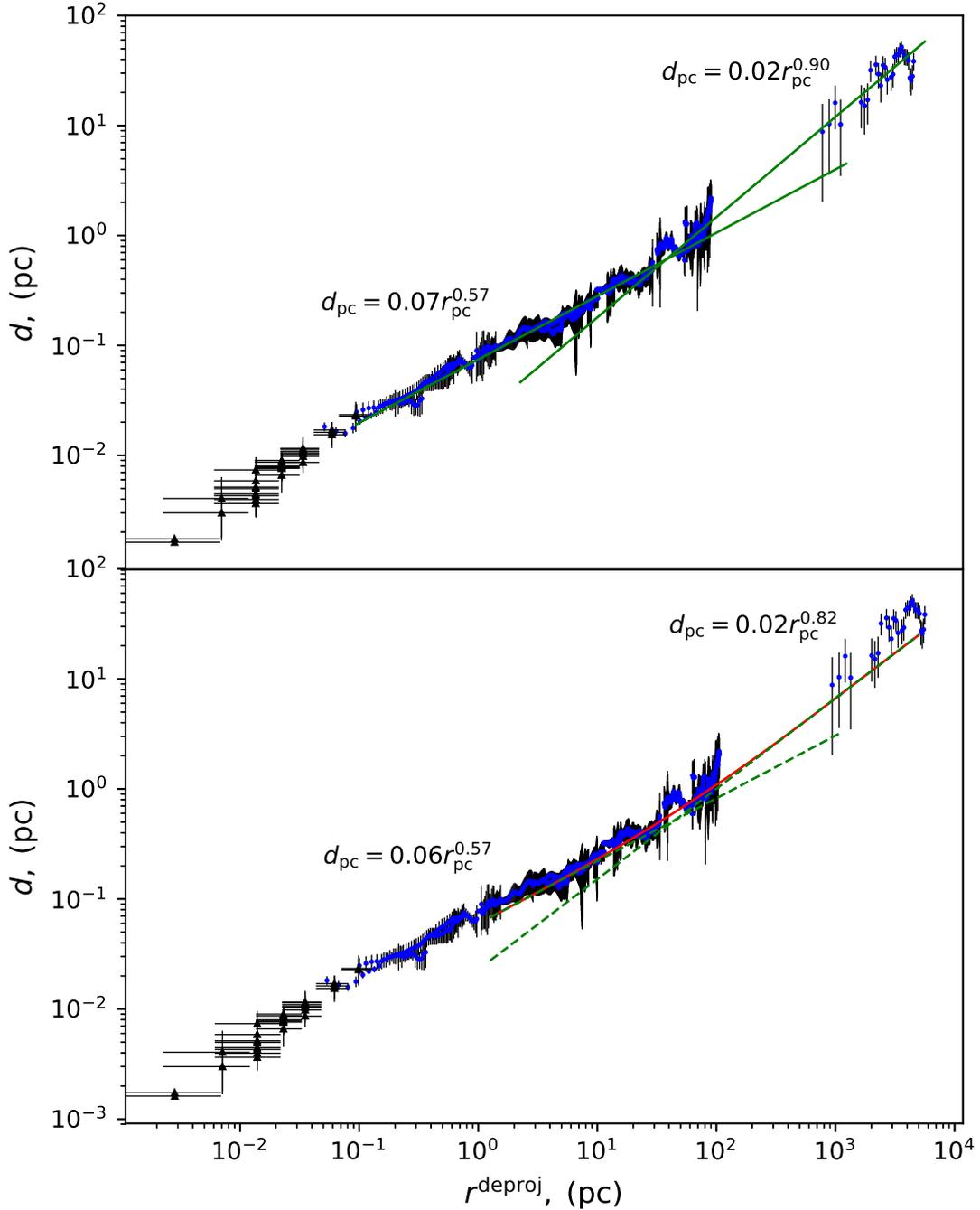}
\caption{The data for M87 jet shape
(blue circles) with error bars (black). The core (black triangles) with error bars (black). Upper plot, two green straight lines --- power-law fits for the observational data.
Lower plot, red line --- 
the model of a jet shape for $b=2.07$ and $\sigma_{\rm M}=20$. Lower plot, two green dashed straight lines --- power-law fits for the model jet boundary shape. Green solid lines and break point in the model on the red curve (the closest point to the green dashed lines intersection) intersect in one point, which allows us to associate the model with the observations. The data and model fits in the parabolic domain approximately coincide, while the fits in conical domain do not, which may be observed at the far right.}
\label{Trial}
\end{figure*}

In the available data set, the cabing position depends on our choice of attributing the data points from 2.3~GHz sample to parabolic or conical domains.
We calculated the cabing position for every point from 2.3-sample being an expected boundary between parabolic and conical domains. We observe that, starting from our ``best'' choice (77 of 394), the cabing position stays more or less constant until about 250 of 394 points of 2.3~GHz sample are attributed to the parabolic domain. We calculate the mean and standard deviation for this ``plato'' sample of cabing positions and jet half-widths. The result is: $d_{break}=0.62\pm 0.02$~pc and $r_{break}=44.7\pm 1.9$~pc.

\begin{table*}
 \begin{minipage}{175mm}
  \caption{Model and derived jet parameters. }
  \label{tableMass}
  \begin{tabular}{llllllllll}
	\hline
  $\sigma_{\rm M}$ & $b$ & $\Gamma\theta_j$ & $\Gamma_{\rm max}$ & $M$ & $a_*$ & $\Psi_0$ & $W_j$ & $R_{\rm L}$ & $R_{\rm L}$ \\
   & & & & $\left(10^9\,M_{\odot}\right)$ & & ($10^{33}\;{\rm G\,cm^2}$) & ($10^{42}$ erg/s) & (pc) & ($r_{\rm g}$) \\
     (1) & (2)  & (3)  & (4)& (5) & (6)  & (7) & (8) & (9) & (10) \\
\hline
5 & 2.045 & $(0.103;\,0.024)$ & 3.4 & $7.7\pm 2.7$ & $0.17\pm 0.06$ & 2.9 & 1.0 & 0.018 & 47\\
10 & 2.050 & $(0.127;\,0.033)$  & 5.1 & $6.6\pm 2.1$ & $0.22\pm 0.06$ & 2.8 & 2.3 &  0.012 & 35 \\ 
20 & 2.070 & $(0.179;\,0.057)$  & 8.6 & $5.2\pm 1.5$  & $0.26\pm 0.07$ & 2.9 & 5.7 & 0.008 & 32 \\
\hline
\end{tabular}\\
\textit{Notes.} Columns are as follows: (1) Michel's magnetisation parameter; (2) exponent in pressure profile; (3) the interval for values of $\Gamma\theta_j$ provided by numerical modelling; (4) maximum Lorentz factor, predicted by our model; (5) estimated BH mass; (6) estimated BH spin; (7) total magnetic flux;
(8) total jet power, associated with the magnetic flux; (9) light cylinder radius in pc; (10) light cylinder radius in $r_{\rm g}$ corresponding to the mass in the column (5).\\ 
\end{minipage}
\end{table*}

We have performed bootstrapping to model how errors (see \autoref{Trial}) in a jet half-width determination affect the cabing position for the ``best'' choice of dividing the sample into parabolic / conical domains. The mean values with standard deviations are $d_{break}=0.61\pm 0.02$~pc and $r_{break}=44.5\pm 1.9$~pc.

We conclude that the typical errors arising from VLBI and MERLIN data are $\pm 0.02$~pc for $d_{break}$ and $\pm 1.9$~pc for $r_{break}$.

The errors provided by \citet{DiMat-03} give the error in the total expression for mass
around 1\% due to errors in particle number density $n=0.170\pm 0.003\;{\rm cm^{-3}}$ and temperature
$kT=0.80\pm 0.01\;{\rm keV}$ measurements. 

However, the scatter in numerical values for $\Gamma\theta_j$ make the major contribution into the error budget, being finally of about an order higher than the errors due do $r_{break}$ and $d_{break}$ position modelling and the errors in measurements of $C_{\dot{M}}$. Thus, we present the result for BH mass as an interval of values corresponding to obtained by our modelling interval for $\Gamma\theta_j$ with the errors {\it (i)--(iii)}. Also we directly give the mean value for BH mass, with the total error including the uncertainty due to $\Gamma\theta_j$ values. See the next Section.

\section{M87 black hole mass}
\label{s:BHM}

We use the closest to the central BH pressure and density measurements by \citet{RF15}: $kT=0.91$~keV and $n_e=0.31\;{\rm cm^{-3}}$ at approximately $r_0=0.22$~kpc. This gives the pressure amplitude $P_0=0.45\times 10^{-9}\;{\rm dyn/cm^2}$ at $r_0=0.22$~kpc from the BH.

The Bondi mass accretion rate was obtained by
\citet{DiMat-03}
basing on measurements of density and temperature of a hot interstellar medium (ISM) using the observed X-ray emission at distances $\lesssim 100$~pc from the black hole. We use the \autoref{Mdot} with the obtained numerical value $C_{\dot{M}}=7\times10^{23}\;{\rm g/s}$. 


We choose the value of an exponent $b$, defining the pressure profile, so as to fit the observed jet shape. For example, for $b=2.07$ and $\sigma_{\rm M}=20$ the model predicts $d\propto r^{0.57}$ for a parabolic domain and $d\propto r^{0.82}$ for a conical domain. In fact, we do not fit precisely both power--laws, describing the observational data in \autoref{s:break_pos}. We set the exponent $b$ so as to fit the parabolic domain (see \autoref{Trial}). In this case, the conical domain of our model still fits the data within the error bars. 

We also need to choose the initial magnetization parameter $\sigma_{\rm M}$. The magnetization $\sigma_{\rm M}$ defines the maximum bulk flow Lorentz factor, that can be achieved by the flow if all the Poynting flux energy is converted to the bulk plasma kinetic energy. It was shown (see, e.g., \citealt{Beskin06, TMN09, Kom09, Lyu09}), that the plasma in highly collimated outflows accelerates effectively only up to approximately $\Gamma\sim\sigma_{\rm M}/2$. Further downstream, the acceleration continues very slowly. Thus, the observed Lorentz factors in M87 can provide us with the estimate for $\sigma_{\rm M}$. \citet{Mertens16} obtained the Lorentz factors of the order of $\Gamma\sim 3$ at $r\sim$ a few parsecs. The detected by \citet{Bir99} Lorentz factors at few hundred parsec is $\sim 10$. We present here results for three values of the magnetization parameter $\sigma_{\rm M}$: 5,10, and 20, which are consistent with the discussed above observed bulk flow Lorentz factors. For these three models we calculate the predicted jet shape profile and find the cabing point position. 

There are theoretical as well as observational constraints on $\Gamma\theta_j$. It was discovered by \citet{TMN09} and \citet{Kom09} that the condition $\Gamma\theta_j<1$ corresponds to the casual connectivity across a jet, ensuring the effective plasma acceleration up to equipartition. \citet{Kom09} showed that the approximate equality $\Gamma\theta_j\approx 1$ should hold for the power-law acceleration regime in a jet. On the other hand, the observations provide the median value $\Gamma\theta_j=0.17$ \citep{MOJAVE_XIV}.
The high resolution data obtained by \citet{Mertens16}
also allows to estimate this value for M87 jet specifically. Measurements and analysis by \citet{Mertens16} provide the apparent opening angle $\theta_{\rm app}$ at the distances $\sim 0.3-4.0$~pc varying from about $18^{\circ}$ closer to the BH to $\approx 7^{\circ}$ further downstream. The intrinsic opening angle depends on the apparent opening angle as $\theta_j=\theta_{\rm app}\sin\varphi/2$, where we use the same viewing angle $\varphi=14^{\circ}$ \citep{Nakamura+18} as was used for the de-projection for the result self-consistency. It gives $\theta_j\approx 0.038$ at $r=0.3$~pc and $0.015$ at $r=4.0$~pc. The Lorentz factor at the same scales varies \citep{Mertens16} from roughly $1.2$ to $\approx 3$. This provides $\Gamma\theta_j\sim 0.046$ at $r=0.3$~pc and $\Gamma\theta_j\sim 0.044$ at $4$~pc, the resultant value being much smaller than theoretical upper boundary for this value. In this paper we use the results of our modelling of a jet structure to bound the possible values of $\Gamma\theta_j$. We calculate the maximum Lorentz factor across a jet and the jet shape boundary derivative $d/r=\tan\theta_j$ for each $r$. We observe that the parameter $\Gamma\theta_j$ does not stay constant along the jet (as was first observed by \citet{Kom09}). It starts at the value $\approx 0.1$ in the parabolic region and runs down up to approximately the cabing point, where it assumes a constant value, corresponding to the maximal Lorentz factor, attained by the jet for the given magnetization, multiplied by the roughly constant opening angle of conical domain.
We use the interval of values for this parameter in our \autoref{MBH} for the BH mass determination. The scatter in this parameter provides the major contribution into errors in the final result for the BH mass.
We should note that $\Gamma\theta_j$, obtained within our modelling is consistent with the result by \citep{Mertens16}, but differs strongly from the assumption $\Gamma\theta_j=1$ by \citet{ZCST14}.


The model parameters and results are presented in \autoref{tableMass}. The model parameters that we set are in columns (1)--(2): the initial jet magnetization and the exponent $b$ set to fit exactly the parabolic domain jet boundary shape. The calculated parameters of the central BH and jet are in columns (3)--(10). We calculate within our model non-dimensional jet shape break parameters $d_*$ and $\tilde{p}_*$ (see \autoref{tableMod}), the interval for $\Gamma\theta_j$, and the maximum Lorentz factor of bulk motion, attained by the flow. We calculate the BH mass using the \autoref{MBH}. We use the measured values for $C_{\dot{M}}$, $P_0$, and $r_0$. We use the results of our fitting the observational data for $d_{break}$ and $r_{break}$ (see \autoref{s:break_pos}), which are consistent with the results by \citet{Asada12}. We also put the model parameters $\sigma_{\rm M}$, $d_*$, $\tilde{p}_*$, and our estimates for $\Gamma\theta_j$. For the result for the SMBH mass in \autoref{tableMass} we present the median value, obtained for each magnetization for the interval $\Gamma\theta_j$ with an error due to uncertainty in this parameter. The same is for the BH spin $a_*$, which we find using \autoref{astar}. The result for the total magnetic flux obtained using \autoref{rbreak} depends on the model parameters and pressure measurements only. 
To calculate the total jet power we use the expression~\citep{Nokhrina18}
\begin{equation}
W_{\rm tot} = \frac{c}{8}\left(\frac{\Psi_0}{\pi R_{\rm L}}\right)^2,
\label{Wtot}
\end{equation}
that relates jet power $W_j$ with the Poynting flux power at the jet base. This formula neglects the initial power in plasma kinetic energy, which is justified for sufficient magnetizations.

We obtain the different values of BH mass for different magnetizations:\\
$M\in(5.0\pm 0.3;\;10.4\pm 0.6)\times 10^{9}\,M_{\odot}$ for $\sigma_{\rm M}=5$; \\ $M\in(4.4\pm 0.3;\;8.7\pm 0.5)\times 10^{9}\,M_{\odot}$ for $\sigma_{\rm M}=10$; \\ 
$M\in(3.8\pm 0.2;\;6.7\pm 0.4)\times 10^{9}\,M_{\odot}$ for $\sigma_{\rm M}=20$. \\
Here the mass interval is due to calculated interval for $\Gamma\theta_j$, and errors are due to errors in jet half-width determination, errors in determination of the cabing point, and errors in mass accretion rate estimate.
The corresponding intervals for the BH spin:\\
$a_*\in(0.11\pm 0.01;\;0.22\pm 0.01)$ for $\sigma_{\rm M}=5$; \\ 
$a_*\in(0.15\pm 0.01;\;0.28\pm 0.02)$ for $\sigma_{\rm M}=10$; \\
$a_*\in(0.19\pm 0.01;\;0.33\pm 0.02)$ for $\sigma_{\rm M}=20$. \\


\section{Results and discussion}

Within the jet model with an electric current locked inside a jet \citep{BCKN-17} we obtained a clear
transition from parabolic to conical jet boundary shape for the ambient pressure given by \autoref{PBondi}.
The break in a jet form occurs as the flow transits from magnetically dominated regime to the rough equipartition between plasma bulk motion kinetic and
magnetic field energy. We propose to associate the positions of a model break with the observed one to 
obtain the jet and BH parameters.
Together with an assumption for the dynamically important magnetic field near the BH and 
measurements of a pressure amplitude, mass accretion rate, and kinematics, we are able
to estimate the BH mass, spin, and total magnetic flux in a jet.

In order to obtain the jet shape break position we use the VLBA and MERLIN data collected by \citet{Nakamura+18}. We should note that not all data from that paper were used here to obtain the result. First of all, our aim is to pin the cabing point position, so the data that is as uniform as possible and that represents the jet boundary shape is of the most importance. We exclude the core data due to errors in core position \citep{Hada11}. We also do not use the European Very Long Base Interferometry Network (EVN) data, because these data cover mainly the HST-1 complex around $r\sim 100$~pc, and may not reflect the jet shape behaviour, but rather the special features of HST-1 itself.

The resultant mass depends on the initial magnetization $\sigma_{\rm M}$. 
For all the values of $\sigma_{\rm M}$, the obtained mass is much bigger than that obtained by \citet{Walsh+13} using a spectral analysis of gas velocity dispersion.
For the high magnetization $\sigma_{\rm M}$, it is somewhat consistent with the result by \citet{Geb11}, based on both gas velocity dispersion measurements and stellar dynamics.
The median value for $\sigma_{\rm M}=10$ is close to the result of \citet{Oldham+16}, obtained basing on the analysis of stellar and cluster dynamics, although the scatter in mass in our paper is bigger due to the strong dependence of the result on the parameter $\Gamma\theta_j$. But the magnetization $=5$, favoured by the kinematics detected in \citep{Mertens16},
points to even bigger BH mass value.

The total magnetic flux depends very weakly on $\sigma_{\rm M}$, and its value of the order of $10^{33}\;{\rm G\,cm^2}$ is consistent with the results by \citet{Nokhrina18}. In contrast, the BH spin depends on $\sigma_{\rm M}$, as the light cylinder radius $R_{\rm L}$ depends on $d_*$ only. This gives the scatter in $a_*$ estimate from $0.11$ to $0.33$ for different magnetizations. Thus, we have obtained the moderate spin parameter of the order of $0.1-0.3$ for M87 SMBH. The numerical simulations \citep{Tchekhovskoy_11, McKinney12} favour the spin $>0.5$ in order to obtain the jet power of the order of $\dot{M}c^2$. On the other hand, semi-analytical and numerical modelling of a BH spin evolution \citep{King-08, Barausse12, VolSik-13, Sesana14} predict moderate  spins $a_*\in(0.1,\,0.7)$ for low redshift $z<2$ AGN, with BH residing in elliptical galaxies tending to have smaller spins, which is consistent with our result.

The result for the total jet power, obtained with \autoref{Wtot}, is consistent with the estimates of an average jet power $W_{j,av}\sim 3\times 10^{42}\;{\rm erg/s}$, needed to evacuate the inner cavities \citep{Young02}. This result is also marginally consistent with the jet power
obtained by \citet{GL17} within a model of the recollimation shock in HST-1 due to a jet interaction with a disk outflow. However, the other theoretical modelling by \citet{Stawarz06} predict higher jet power $\sim 10^{43}-10^{44}\;{\rm erg/s}$  needed to feed the radio lobes \citep{Owen_00}.
This may be an indication that the rough estimate of jet power by \citet{Beskin10} without a numerical factor
is more robust, providing for M87 total jet power a few of $10^{43}\;{\rm erg/s}$. Indeed, the factor $1/8$ does provide the correlation of a magnetic flux with the averaged over large period of time power \citep{Nokhrina18}, but it also depends on the particular choice of MHD integrals. 

We are able to fit the the observed jet boundary shape with the theoretical curve in the parabolic domain with $b\approx 2.05-2.07$, which is consistent with Bondi accretion flow models \citet{QN00, NF11}. However, the direct measurements of a particle number density in ISM either by X-ray observations \citep{RF15}, or by modelling the Faraday rotation measure on the ambient medium \citep{Park-19}, provide $n\propto r^{-1}$, which corresponds to smaller $b$ for adiabatic flow. This caveat may be solved if the temperature rises closer to the central source, as was predicted by \citet{QN00} and discussed in \citet{RF15}.

Our model means that the more or less effective plasma acceleration takes place up to the cabing point, or, approximately, up to HST-1 \citep{Asada12}. This is different from modelling by \citet{Mertens16}, in which the acceleration saturation is set at the distance approximately $4$~pc from the jet base. However, the longer acceleration domain obtained within our model is consistent with the observed by~\citet{Bir99} Lorentz factors of the order of $10$ at the HST-1 (the cabing region \citep{Asada12}). The kinematics obtained by radio interferometric measurements \citep{Mertens16, Lister-19} with low detected Lorentz factors favour the smaller magnetizations $\sigma_{\rm M}=5\div 10$, with the predicted bigger central BH mass. On the other hand, the optical observations of velocities at $\sim 100$~pc scales by \citet{Bir99} favour $\sigma_{\rm M}=20$.

As was stressed, the method for determining BH mass proposed above is based on the existence of a statistical dependence \autoref{PsiMAD}, which relates the total magnetic flux $\Psi_{0}$ to the accretion rate ${\dot M}$. In cases where the accretion rate ${\dot M}$ can be found independently, the procedure for determining the mass may be changed. In particular,
note that the relation \autoref{Mdot} provides the accretion rate ${\dot M} \approx 0.2 - 0.4 \, M_{\odot}$ yr$^{-1}$ for the masses we estimated for M87. We plan to address the question of mass determination for the other sources with the detected jet boundary shape break, in particular 1H0323+342 \citep{Hada18}, in the future work.

Using the definition of a magnetization parameter $\sigma_{\rm M}$, 
one can rewrite the mass ejection rate ${\dot M}_{\rm eject}$ in a jet in the form
\begin{equation}
{\dot M}_{\rm eject} \approx \frac{W_{\rm tot}}{\sigma_{\rm M}c^2},
\label{Meject}
\end{equation}
where $W_{\rm tot}$ \autoref{Wtot} is the total energy losses in a jet. This value does not depend on the assumed jet composition. 
For $\sigma_{\rm M} \sim 10$ we obtain the reasonable value of the mass ejection rate ${\dot M}_{\rm eject} \sim 10^{-4} M_{\odot}$ yr$^{-1}$. 
This mass loss rate in a jet corresponds to the mass density $\rho$ through the given
jet cross section $\dot{M}_{\rm eject}=\rho c S$. At the distance 1~pc from the ``central engine'' (typical distance where the particle number density is calculated through the core shift effect) the jet radius $\sim 0.1$~pc. If the mass density
is defined by the electrons, than the particle number density at 1~pc is of the order of 100 ${\rm cm}^{-3}$, which is in agreement with another independent evaluation by core-shift data~\citet{NBKZ15}. For the protons the appropriate particle number density is 1800 times less. Therefore, this result points at the mainly electron-positron composition of the M87 jet.

\subsection{Corroboration by the EHT results}
\label{ss:corr}

The brand new EHT results \citep{EHT_I} provide the BH mass in M87 as $M=(6.5\pm 0.7)\times 10^9\;{\rm M}_{\odot}$. This result is in agreement with the choice of $\sigma_{\rm M}=10,\;20$, in accordance with the observed kinematics \citep{Bir99, Mertens16}. 

The BH mass value $M=6.5\times 10^9\;{\rm M}_{\odot}$
corresponds to the BH and jet properties, listed in \autoref{tableEHTcorr}.
\begin{table}
  \caption{Predicted jet and BH parameters for $M=6.5\times 10^9\;{\rm M}_{\odot}$.\label{tableEHTcorr}}
 \begin{tabular}{ccccc}
  \hline
  $\sigma_{\rm M}$ & $\Gamma\theta_j$ & $\phi$ & $a_*$ & $W_j$ \\
  & & & & ($10^{42}$ erg/s) \\
     (1) & (2) & (3) & (4) & (5) \\
\hline
10 & $0.059$ & $3.1$ & $0.21$ & $2.3$ \\
20 & $0.060$ & $3.1$ & $0.32$ & $5.7$ \\
\hline 
\end{tabular}
\end{table}
We predict that the jet is highly casually connected $\Gamma\theta_j\ll 1$. The disk state is far from the MAD ($\phi\sim 50$), and obtained value $\phi\approx 3$ (in Gaussian units) suggests the SANE disk state. We also predict the moderate spin of the order of $0.2-0.3$. This value has not been probed by the \citep{EHT_V} modelling. The total jet power corresponds to the highest obtained by the EHT collaboration results, being closer to the estimates obtained in the previous works \citep{Stawarz06, Owen_00, Young02, GL17}. Again, this power may be higher by the factor of about four, but this needs further investigation. 

The proposed method of estimating the BH mass and spin, total magnetic flux in a jet, and total jet power, may prove to be a powerful instrument in probing the BH physics. It is in full accordance with the EHT results and multitude of previous studies of BH environment, as well as jet morphology and kinematics.
At the same time, this instrument presented here requires resolution of jet boundary shapes on the scale of tens of parsecs, or $10^5$ gravitational radii, which is an attainable goal for the modern VLBI systems. We also note that the comparison of the ``traditional'' cm--dm--wavelength VLBI results discussed in this work and EHT results on M87 will offer a powerful calibration method for future interpretation of high-resolution studies in many AGN.



\section*{Acknowledgements}
We thank the anonymous referee for
suggestions which helped to improve the paper. This research was supported in part
by the 5-100 Russian Academic Excellence Project (Agreement number
05.Y09.21.0018) and by the Russian Foundation for Basic Researches (grant 17-02-00788).

\bibliographystyle{mnras}
\bibliography{nee1}
\bsp    
\label{lastpage}

\end{document}